\documentclass{article}

\usepackage{arxiv}

\usepackage[utf8]{inputenc} 
\usepackage[T1]{fontenc}    
\usepackage{hyperref}       
\usepackage{url}            
\usepackage{booktabs}       
\usepackage{amsfonts}       
\usepackage{nicefrac}       
\usepackage{microtype}      
\usepackage{lipsum}		
\usepackage{graphicx}
\usepackage{doi}
\usepackage{amsmath}
\usepackage{cite}

\title{An extended equivalent circuit model for analysis of arrays of sub-wavelength holes perforated in a metallic film}

\author{ \href{https://orcid.org/0000-0002-7455-7990}{\includegraphics[scale=0.06]{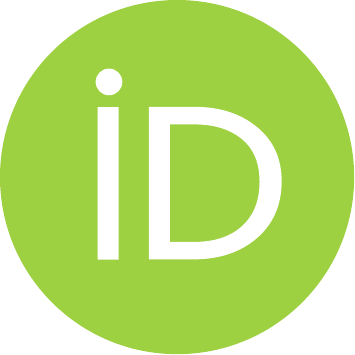}\hspace{1mm}Mohammad Pasdari-Kia} \\
	Department of Electrical Engineering\\
	Sharif University of Technology\\
	Tehran, Iran \\
	\texttt{mohammad.pasdarikia@sharif.edu} \\
	\And
	\href{https://orcid.org/0000-0002-7764-2437}{\includegraphics[scale=0.06]{orcid.pdf}\hspace{1mm}Mohammad Memarian} \\
	Department of Electrical Engineering\\
	Sharif University of Technology\\
	Tehran, Iran \\
	\texttt{mmemarian@sharif.edu} \\
		\And
	\href{https://orcid.org/0000-0003-3479-3830}{\includegraphics[scale=0.06]{orcid.pdf}\hspace{1mm}Amin Khavasi} \\
	Department of Electrical Engineering\\
	Sharif University of Technology\\
	Tehran, Iran \\
	\texttt{khavasi@sharif.edu}
}

\renewcommand{\shorttitle}{Preprint}

\begin{document}
\maketitle

\begin{abstract}
		Periodic arrays of sub-wavelength holes, due to a variety of applications (sensors, polarizers, filters) and unique abilities in manipulating different characteristics of impinging light, have been the subject of many studies in recent years. This paper presents a new high-precision circuit model to investigate these structures' electromagnetic response, which provides higher accuracy than available literature. To develop the model, a multi-mode approach is taken for inside the apertures, and the proposed model is presented in three general forms which are explained in detail. Along with the proposed circuit model, accurate spatial profiles for square and circular apertures are presented in which the transverse components of the electric field are approximated. In this paper, it is shown that structures such as an array of PEC pillars can be analyzed similar to other arrays of apertures, and the circuit model accurately predicts the behavior of these structures. In some practical applications, such as sensors and TCEs (transparent conducting electrodes), the array is embedded in a layered medium, and the circuit model with minor changes and few modifications can analyze these structures with high accuracy.
\end{abstract}


\section{Introduction}
The propagation of electromagnetic waves along periodic arrays of sub-wavelength apertures has been the subject of many studies in the last two decades. These structures have various applications, including sensors \cite{chen2019enhanced,conteduca2021dielectric,yahiaoui2016terahertz,gordon2008new,dhawan2008plasmonic}, filters \cite{panwar2017progress}, absorbers \cite{huang2022broadband}, waveguides \cite{liang2020metallic,ahmadi2023babinet}, lenses \cite{garcia2019plasmonic,huang2018superfocusing}, etc. Along with these novel applications, observation of the physical phenomena such as extraordinary transmission (EoT) \cite{ebbesen1998extraordinary,ruiz2022single} and broadband Brewster transmission \cite{paniagua2016generalized,fan2021optical,edalatipour2015physics} has increased the importance of these structures.

In recent years, understanding and justifying the physical behaviors of the array of holes has been researched using various analytical methods \cite{de2007colloquium,martin2001theory}. For instance, the extraordinary transmission at optical frequencies may be understood by coupling incident light with the surface plasmon polariton modes supported at the boundary between metals and dielectrics \cite{liu2008microscopic}. On the other hand, this enhanced transmission at microwave frequencies through perforated PEC structures, has recently been linked to the role of proper complex modes \cite{ansari2020local}. Providing an analytical method in the form of a circuit model can become a tool for designing and predicting these behaviors. This work focuses on microwave and terahertz frequencies, where the permittivity of the metals becomes high enough to justify perfect electric conductor (PEC) approximation. The transmission line model is considered in this paper due to its appropriate accuracy and high flexibility, which can be easily generalized to multilayer structures.

Modeling a waveguide structure in the form of a transmission line has been researched for a long time and has numerous applications in obtaining scattering parameters \cite{marcuvitz1951waveguide,pozar2011microwave}. The transmission line model is used in a wide frequency range from microwave to infrared \cite{pasdari2022generalized,pasdari2023variational,costa2014overview,zhao2011homogenization,khavasi2014analytical}. With the help of the transmission line, the array of apertures in a PEC film has already been investigated \cite{molero2021cross,borgese2020simple,kaipa2010circuit,medina2008extraordinary}. In the first developed circuit model, the array of holes was analyzed by considering the thickness of the PEC film. In this case, the circuit model was created by considering only the dominant mode inside the apertures \cite{medina2009extraordinary,khavasi2015corrections}.

In previous works, a circuit model has been developed for rectangular apertures by considering only $TE_{01}$ as the dominant mode \cite{khavasi2015corrections,marques2009analytical}. In addition to two-dimensional arrays, one-dimensional arrays of slits have also been investigated by considering the $TEM$ mode as the dominant mode using the circuit model \cite{medina2009extraordinary,yarmoghaddam2014circuit}. If the thickness of the aperture is large enough that the excited high-order modes inside the hole are damped from the first opening to the second opening of the hole, the model has high accuracy.
In these models, if the structure's thickness decreases, the error increases due to high-order modes; additionally, the error increases more rapidly for circular-shape apertures compared to rectangular apertures. Due to the widespread applications of metasurfaces and frequency selective surfaces, we need to create models for thin arrays.

Frequency selective surfaces are thin periodic surfaces that are important both in terms of industrial applications as well as in terms of fundamental science. The array of apertures perforated in a PEC thin film is a kind of FSS, and many powerful circuit models have been proposed for these structures that are suitable for better design and understanding of the behavior of these structures \cite{mesa2018efficient,alex2021exploring}. In these structures, the circuit model is created with the help of the spatial profile of the transverse electric field at the opening of the aperture. Previous works have provided profiles for rectangular \cite{alex2021exploring}, cross \cite{molero2021cross}, and annular holes \cite{rodriguez2017annular,alex2021exploring}. Initially, the circuit model was developed for a structure consisting of a single layer of the intended FSS \cite{costa2014overview}; then, the circuit model was improved for the case where the FSS is embedded in a layered medium \cite{rodriguez2012analytical,rodriguez2015analytical}. Structures with multiple dielectric layers and multiple FSS layers were then examined using circuit models \cite{molero2021cross,alex2021exploring}, but all of these circuit models were developed regardless of the FSS's thickness. It should be noted that effect of thickness can cause a considerable error; however, the thickness is about one-fiftieth of the structure's period.

In this paper, an extended circuit model is developed, which is much faster than full-wave simulations and is much more accurate than its predecessors. The circuit model is formulated in such a way that it takes into account high-order modes inside the holes; thus, it has high accuracy in low thickness structures. A closed-form expression is presented for the spatial profile of the transverse electrical field on the circular aperture which means that, unlike previous models, we do not need to extract the field profile from the full-wave simulators. It is shown that the previous electric field profile for the square hole has a large error when the width of the square hole is large  (the width of the square is greater than half the period) \cite{molero2021cross}, and this problem is well solved by presenting the new profile. Unlike square and circular holes, an array of PEC pillars always supports propagating mode that makes interesting properties \cite{edalatipour2015physics}, and these structures are also analyzed analytically with the help of the proposed circuit model in this work.

The paper is organized as follows: Section 2 is devoted to describing how the proposed circuit model is developed, and three different cases: single-mode, multi-mode, and modified single-mode circuit models are described. In section 3, the low-thickness structures are investigated. In section 4, the proposed circuit model is extended to multilayered structures; finally, a conclusion is described in section 5.

\section{The proposed circuit model}
Periodic structures can be analyzed similar to other waveguide structures by considering a unit cell as a waveguide; thus, the formulation presented can be extended to any waveguide structure, but we focus only on the square array of sub-wavelength holes. In addition to the previous circuit models, environmental models for the desired structure have also been investigated \cite{edalatipour2015physics,edalatipour2012creation}. Similarly, the isotropic/anisotropic environmental model is created only by considering the dominant mode inside the aperture, which means that these methods also have a large error for a thin array of holes. The following formulation is presented by considering all the modes inside the hole.
\begin{figure}[!ht]
	\centering
	\includegraphics[width=13cm,height=5cm]{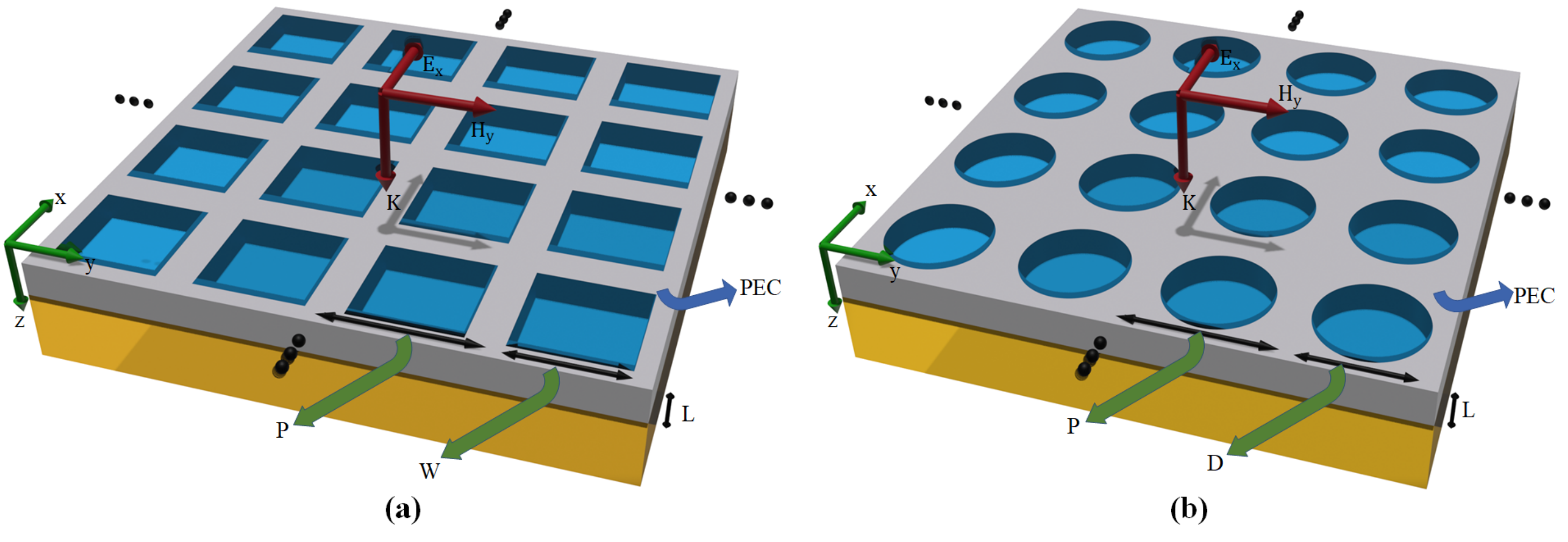}
	\caption{Figure (a) shows the array of perforated square apertures in a PEC film, and (b) shows the array of circular holes. In two structures, the period is P in both directions, the width of the square is W, and the circle's diameter is D. Refractive index of the upper and lower (yellow layer) environments are $n_{1}$ and $n_{3}$, respectively, and the holes are filled with 
		refractive index $n_{2}$.}\label{periodic_array}
\end{figure}

Figure \ref{periodic_array} shows the schematic of the structures which are illuminated by a normal incident plane wave. The upper environment of the structure is region I ($n_{1}$), inside the holes is region II ($n_{2}$), and the lower environment is region III ($n_{3}$). For simplicity’s sake, we first assume the structures depicted in Fig. \ref{periodic_array} is semi-infinite, i.e., $L \rightarrow \infty$.  Impinging of a plane-wave on a semi-infinite structure excites aperture modes in region II, and the Floquet modes in region I. The Floquet expansion of the transverse (x, y components) electric and magnetic field at the discontinuity ($z=0$) can be
written as follow \cite{molero2021cross,rodriguez2015analytical}:
\begin{equation}\label{up_mode_expansion1}
	\begin{cases}
		E^{1}(x,y)=(1 + R)e_{0}^{1}(x,y)  + \sum_{h}^{'}V_{h}^{1} e_{h}^{1}(x,y)
		\vspace{.5cm}\\
		H^{1}(x,y)=(1 - R)Y_{0}^{1}(\hat z  \times e_{0}^{1}(x,y)) - \sum_{h}^{'}V_{h}^{1} Y_{h}^{1}(\hat z \times e_{h}^{1}(x,y))
	\end{cases}
\end{equation}
where the expression $e_{0}^{1}(x,y)$ is the tangential component of the incident wave whose reflection is R. The superscript refers to the region, and the prime in the series of Eq. (\ref{up_mode_expansion1})  indicates that the incident wave is excluded from the summation.  $e_{h}^{1}(x,y)$ is the normalized transverse electric field of $h$th Floquet harmonic ($h$ is associated with a pair of integer numbers $mn$). Similarly, expansion of aperture modes can be written as
\begin{equation}\label{aperture_mode_expansion1}
	\begin{cases}
		E^{2}(x,y)=V_{0}^{2}e_{0}^{2}(x,y)  + \sum_{h}^{'}V_{h}^{2} e_{h}^{2}(x,y)
		\vspace{.5cm}\\
		H^{2}(x,y)=V_{0}^{2}Y_{0}^{2}(\hat z \times e_{0}^{2}(x,y)) + \sum_{h}^{'}V_{h}^{2} Y_{h}^{2}(\hat z \times e_{h}^{2}(x,y))  \hspace{.7cm}
	\end{cases}
\end{equation}
where $e_{0}^{2}(x,y)$ is the tangential component of the dominant mode inside the aperture. It should be noted that we want to use the dominant mode to create the transmission line; thus, we have written this mode separately from the other modes. Similarly, $e_{h}^{2}(x,y)$ is the transverse electric field of harmonic $h$ inside the aperture. Modes inside the aperture can be used in equations depending on the type of aperture but the Floquet modes in region I can be written as \cite{rodriguez2015analytical, molero2021cross}:
\begin{equation}\label{fl1}
	\begin{aligned}
		e_{h}^{1}(x,y)=\frac{e^{-jK_{th}^{1}.\hat{\rho}}}{PP} \hat{e_{h}}^{1} \hspace{1cm} \hat{\rho}=x\hat{x} + y\hat{y}
	\end{aligned}
\end{equation}
\begin{equation}\label{fl2}
	\begin{aligned}
		k_{th}^{1}=k_{xm}^{1}\hat{x} + k_{yn}^{1}\hat{y} = (k_{m}^{1} + k_{x0}^{1} )\hat{x} + (k_{n}^{1} + k_{y0}^{1})\hat{y}
	\end{aligned}
\end{equation}
\begin{equation}\label{fl3}
	\begin{aligned}
		k_{x0}^{1}=k^{1}\sin(\theta)\cos(\phi) \hspace{1cm}    k_{y0}^{1}=k^{1}\sin(\theta)\sin(\phi)
	\end{aligned}
\end{equation}
\begin{equation}\label{fl4}
	\begin{aligned}
		k_{m}^{1}=\frac{2\pi m}{P} \hspace{1cm}  k_{n}^{1}=\frac{2\pi n}{P}
	\end{aligned}
\end{equation}
\begin{equation}\label{fl5}
	\hat{e_{h}}^{1}=
	\begin{cases}
		\hat{k_{th}}^{1}   \hspace{.5cm}       & TM
		\vspace{.5cm}\\
		(\hat{k_{th}}^{1} \times \hat{z} )  \hspace{.5cm}  & TE
	\end{cases}
	\hspace{1cm}  \hat{k_{th}}^{1}=\frac{k_{th}^{1}}{\mid k_{th}^{1} \mid} \hspace{1cm}
\end{equation}
The modal admittances $Y_{h}^{1}$ are given by
\begin{equation}\label{fl6}
	Y_{h}^{1}=\frac{1}{\eta^{1}}
	\begin{cases}
		\frac{k^{1}}{k_{zh}^{1}}   \hspace{.5cm}       & TM
		\vspace{.5cm}\\
		\frac{k_{zh}^{1}}{k^{1}}   \hspace{.5cm}  & TE
	\end{cases}
	\hspace{.5cm}  k_{zh}^{1}=\sqrt{(k^{1})^{2}  -  \mid k_{th}^{1} \mid ^{2}}
\end{equation}
with 
\begin{equation}\label{f21}
	\begin{aligned}
		k^{1}=n_{1}k_{0}  \hspace{1cm} \eta^{1}=\frac{\eta^{0}}{n_{1}}
	\end{aligned}
\end{equation}
where $k_{0}$ is the vacuum wavenumber, $\eta^{0}$ is the intrinsic impedance of free space. These equations are for any incident angle; however, we focus on normal incident waves in this article for simplicity in presentation. In the case of a normal incident wave ($\theta = 0,  \phi = 0$), we can use modes of a waveguide with PEC and PMC walls instead of using the written Floquet modes which decrease computational volume significantly \cite{marques2009analytical,kaipa2010circuit}. Aperture modes can be used in equations depending on the shape of the aperture. In this paper, square and circular aperture modes are used (the dominant modes are $TE_{01}$ and $TE_{11}$ for the square and circular aperture, respectively).

To create the circuit model, we use the transverse electric field at the opening of the aperture ($E_{a}$). The transverse electric field on the aperture can be written as $E_{a}(x,y)=f(\omega)e_{a}(x,y)$. By changing the frequency, almost the spatial profile of the transverse electric field on the aperture doesn't change ($e_{a}(x,y)$), and only its factor changes ($f(\omega)$). This assumption is the basis for creating many analytical models that were referred to in the introduction. To develop the circuit model, we must apply boundary conditions at the discontinuity. The boundary conditions of the tangential electric fields at the discontinuity are
\begin{equation}\label{eq_Exa1}
	\begin{aligned}
		(1 + R)e_{0}^{1}(x,y)  +  \sum_{h}\phantom{}^{'}V_{h}^{1} e_{h}^{1}(x,y)=E_{a}(x,y)
	\end{aligned}
\end{equation}
\begin{equation}\label{eq_Exa2}
	\begin{aligned}
		V_{0}^{2}e_{0}^{2}(x,y)  + \sum_{h}\phantom{}^{'}V_{h}^{2} e_{h}^{2}(x,y)=E_{a}(x,y)
	\end{aligned}
\end{equation}

We can write the orthogonality of modes as
\begin{equation}\label{ortho}
	\iint\limits_{c} (e_{h}^{1} \times (\hat{z}\times e_{k}^{1})^{*}).\hat{z} ds =0  \hspace{.5cm} if \hspace{.5cm} k\neq h
\end{equation}

If we use the orthogonality of Floquet modes in Eq. (\ref{eq_Exa1}), the coefficients of Floquet harmonics are calculated as:
\begin{equation}\label{orth_e1}
	\begin{cases}
		(1 + R)=\frac{\iint\limits_{a} (E_{a}(x,y) \times (\hat{z}\times e_{0}^{1})^{*}).\hat{z} ds}{\iint\limits_{c} (e_{0}^{1} \times (\hat{z}\times e_{0}^{1})^{*}).\hat{z} ds}=\frac{A_{0}}{C_{0}}
		\vspace{.5cm}\\
		V_{h}^{1}=\frac{\iint\limits_{a} (E_{a}(x,y) \times (\hat{z}\times e_{h}^{1})^{*}).\hat{z} ds}{\iint\limits_{c} (e_{h}^{1} \times (\hat{z}\times e_{h}^{1})^{*}).\hat{z} ds}=\frac{A_{h}}{C_{h}}
		\vspace{.5cm}\\
		V_{h}^{1}=(1 + R)\frac{A_{h}C_{0}}{C_{h}A_{0}}
	\end{cases}
\end{equation}

Similarly, using the orthogonality of aperture modes and Eq. (\ref{eq_Exa2}) implies that:
\begin{equation}
	\begin{cases}
		V_{0}^{2}=\frac{\iint\limits_{a} (E_{a}(x,y) \times (\hat{z}\times e_{0}^{2})^{*}).\hat{z} ds}{\iint\limits_{a} (e_{0}^{2} \times (\hat{z}\times e_{0}^{2})^{*}).\hat{z} ds}=\frac{B_{0}}{D_{0}}
		\vspace{.5cm}\\
		V_{h}^{2}=\frac{\iint\limits_{a} (E_{a}(x,y) \times (\hat{z}\times e_{h}^{2})^{*}).\hat{z} ds}{\iint\limits_{a} (e_{h}^{2} \times (\hat{z}\times e_{h}^{2})^{*}).\hat{z} ds}=\frac{B_{h}}{D_{h}}
		\vspace{.5cm}\\
		V_{h}^{2}=V_{0}^{2}\frac{B_{h}D_{0}}{D_{h}B_{0}}
	\end{cases}
\end{equation}

The continuity of the tangential magnetic field on the aperture's surface helps us to create the circuit model. The continuity of the magnetic field on the aperture can be written as:
\begin{equation}\label{H_con}
	\begin{aligned}
		(1 - R)Y_{0}^{1}(\hat z \times e_{0}^{1}(x,y))  - \sum_{h}\phantom{}^{'}V_{h}^{1} Y_{h}^{1}(\hat z \times e_{h}^{1}(x,y))= 
		V_{0}^{2}Y_{0}^{2}(\hat z \times e_{0}^{2}(x,y))  +  \sum_{h}\phantom{}^{'}V_{h}^{2} Y_{h}^{2}(\hat z \times e_{h}^{2}(x,y))
	\end{aligned}
\end{equation}

Multiplying both sides of Eq. (\ref{H_con}) by $E_{a}(x,y)$ and a few mathematical operations yields
\begin{equation}\label{H_con_2}
	\begin{aligned}
		Y_{0}^{1}(1 - R)A_{0}^{*}  - \sum_{h}\phantom{}^{'}Y_{h}^{1} V_{h}^{1}A_{h}^{*} =V_{0}^{2}Y_{0}^{2} B_{0}^*  +  \sum_{h}\phantom{}^{'}Y_{h}^{2}V_{h}^{2} B_{h}^{*}
	\end{aligned}
\end{equation}

If we now place $V_{h}^{1}$ and $V_{h}^{2}$ in Eq. (\ref{H_con_2}), the following equation is obtained.
\begin{equation}\label{circuit_eq}
	\begin{aligned}
		Y_{0}^{1}\frac{1 - R}{1 + R} =Y_{0}^{2} \frac{C_{0}B_{0}B_{0}^*}{D_{0}A_{0}A_{0}^*} +\sum_{h}\phantom{}^{'}Y_{h}^{2} \frac{B_{h}B_{h}^{*}C_{0}}{A_{0}A_{0}^{*}D_{h}}  +\sum_{h}\phantom{}^{'} Y_{h}^{1} \frac{A_{h}A_{h}^{*}C_{0}}{A_{0}A_{0}^{*}C_{h}}
	\end{aligned}
\end{equation}

\begin{figure}[ht]
	\centering
	\includegraphics[width=4.5cm,height=3.5cm]{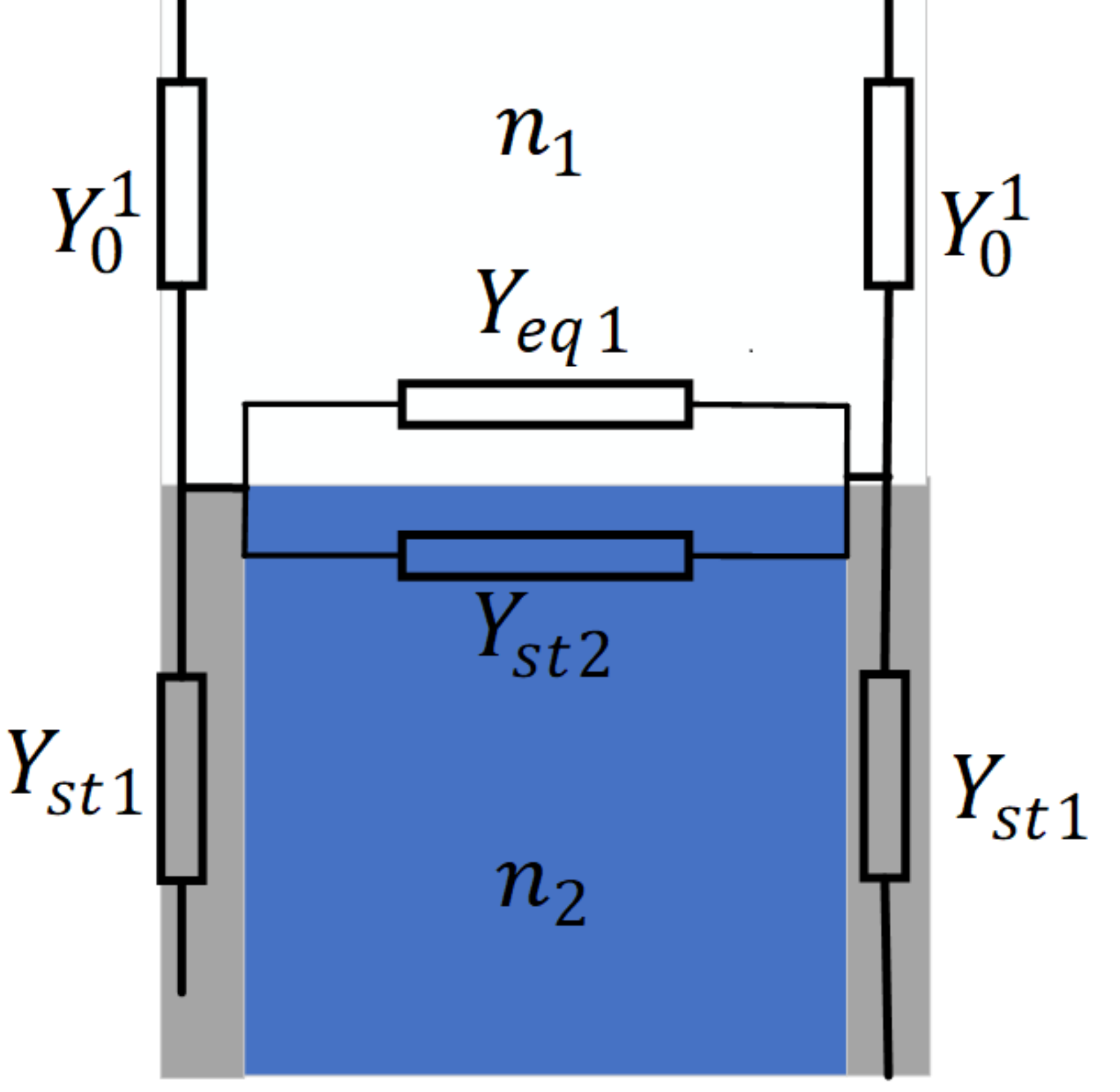}
	\caption{Circuit model for a semi-infinite array of sub-wavelength apertures.
		\label{fig:circuit_1}
	}
\end{figure}

Equation \ref{circuit_eq} can be converted to the circuit model shown in
Fig. \ref{fig:circuit_1}, where the circuit model admittances are in the form of equations in Eq. (\ref{circuit_parameter1}). In the obtained admittances, the frequency dependence of $E_{a}(x,y)$ is eliminated; Thus, the admittances depend only on the spatial profile of the transverse electric field at the opening of the aperture and the electric field value on the aperture does not affect the admittances. In this circuit model, $Y_{eq1}$ is the result of all the Floquet modes except the incident wave in region I (the incident mode is used as the transmission line $Y_{0}^{1}$). $Y_{st1}$ is the result of the dominant mode inside the aperture, and the effect of other modes inside the aperture is modeled as admittance $Y_{st2}$.
\begin{equation}\label{circuit_parameter1}
	\begin{cases}
		Y_{eq1}=\sum_{h}^{'} Y_{h}^{1} \frac{A_{h}A_{h}^{*}C_{0}}{A_{0}A_{0}^{*}C_{h}}
		\vspace{.5cm} \\
		Y_{st1}=Y_{0}^{2} \frac{C_{0}B_{0}B_{0}^*}{D_{0}A_{0}A_{0}^*}
		\vspace{.5cm} \\
		Y_{st2}=\sum_{h}^{'}Y_{h}^{2} \frac{B_{h}B_{h}^{*}C_{0}}{A_{0}A_{0}^{*}D_{h}} \hspace{1.5cm}
	\end{cases}
\end{equation}

The circuit model shown in Fig. \ref{fig:circuit_1}  is for semi-infinite structures. In order to develop this model in a way that is compatible with structures illustrated in Fig. \ref{periodic_array}, the model must be converted to Fig. \ref{fig:circuit_3} and modification on its impedances should also be considered. It should be noted that the evanescent modes are equivalent to the energy storage elements, so the TE modes are equivalent to the inductors, and the TM modes are equivalent to the capacitors. The corresponding admittance of each harmonic is proportional to the energy stored in that mode. Unlike the semi-infinite structure, the aperture modes do not extend to infinity in the finite-thickness array; therefore, we must modify the admittances related to aperture modes. We use the concept of energy to modify the admittances. Admittances related to aperture modes to represent the energy stored in the finite thickness must be multiplied by $(1 - e^{j2K_{zh}^{2}L})$. Admittance related to region I ($Y_{eq1}$) does not change; similarly, admittance related to region III is written. It should be noted that the dominant mode inside the hole does not need to be modified because it has become a transmission line. In fact, the dominant mode plays the main role in energy transfer from the upper opening of the hole to the lower opening. Additionally, the dominant mode is the first mode that converts from the evanescent mode to the propagating mode as the frequency increases, therefore transmission line is a better choice for the dominant mode. According to the above, the admittances of a finite-thickness array of holes in Fig. \ref{periodic_array} are written as:
\begin{figure}[ht]
	\centering
	\includegraphics[width=5cm,height=4.5cm]{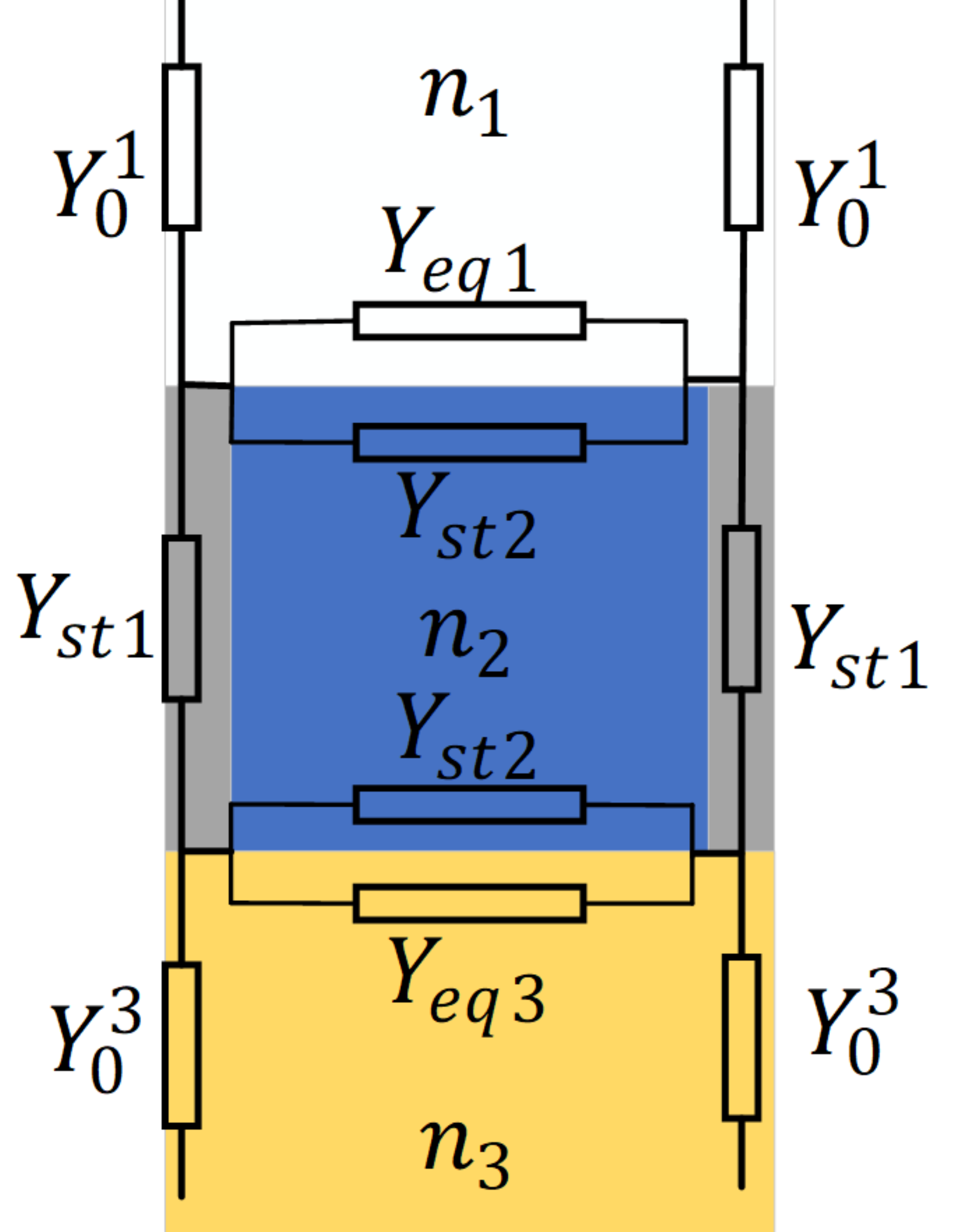}
	\caption{The final circuit model for finite-thickness arrays of sub-wavelength holes shown in Fig. \ref{periodic_array}.}\label{fig:circuit_3}
\end{figure}

\begin{equation}\label{circuit_model_3}
	\begin{cases}
		Y_{eq1}=\sum_{h}^{'} Y_{h}^{1} \frac{A_{h}A_{h}^{*}C_{0}}{A_{0}A_{0}^{*}C_{h}}
		\vspace{.5cm} \\
		Y_{st1}=Y_{0}^{2} \frac{C_{0}B_{0}B_{0}^*}{D_{0}A_{0}A_{0}^*}
		\vspace{.5cm} \\
		Y_{st2}=\sum_{h}^{'}Y_{h}^{2} \frac{B_{h}B_{h}^{*}C_{0}}{A_{0}A_{0}^{*}D_{h}}(1 - e^{j2K_{zh}^{2}L})
		\vspace{.5cm} \\
		Y_{eq3}=\sum_{h}^{'} Y_{h}^{3} \frac{A_{h}A_{h}^{*}C_{0}}{A_{0}A_{0}^{*}C_{h}}
	\end{cases}
\end{equation}

The bottom opening of the aperture has the same profile as the top opening of the aperture, so in the bottom opening of the aperture, the admittance $Y_{st2}$ will be unchanged. Only the directions of the aperture modes at the top and bottom are opposite, which does not make a difference in the value of this admittance. Due to the similarity of the profiles at the top and bottom openings, $Y_{eq3}$ is computed like $Y_{eq1}$. The final circuit model is examined in 3 forms which are described below.

\subsection{Single-Mode Circuit Model (SMCM)}
This case has been discussed in previous works \cite{khavasi2015corrections,marques2009analytical,medina2009extraordinary}, but additional explanations are provided to clarify the differences between the cases in this subsection. The SMCM considers all the Floquet modes in regions I and III, but it considers only the dominant mode inside the aperture ($Y_{st2}=0$). The formulation in previous articles has been written in such a way that the electric field profile at the aperture opening is considered the same as the electric field profile of the dominant mode. For instance, in square and rectangular holes, the electric field profile on the aperture is considered as follows:
\begin{equation}\label{TE01_model}
	E_{a}(x,y)=e_{0}^{2}(x,y)=\cos(\frac{\pi y}{W}) \hat{x}
\end{equation}
\begin{figure}[!ht]
	\centering
	\includegraphics[width=8cm,height=6cm]{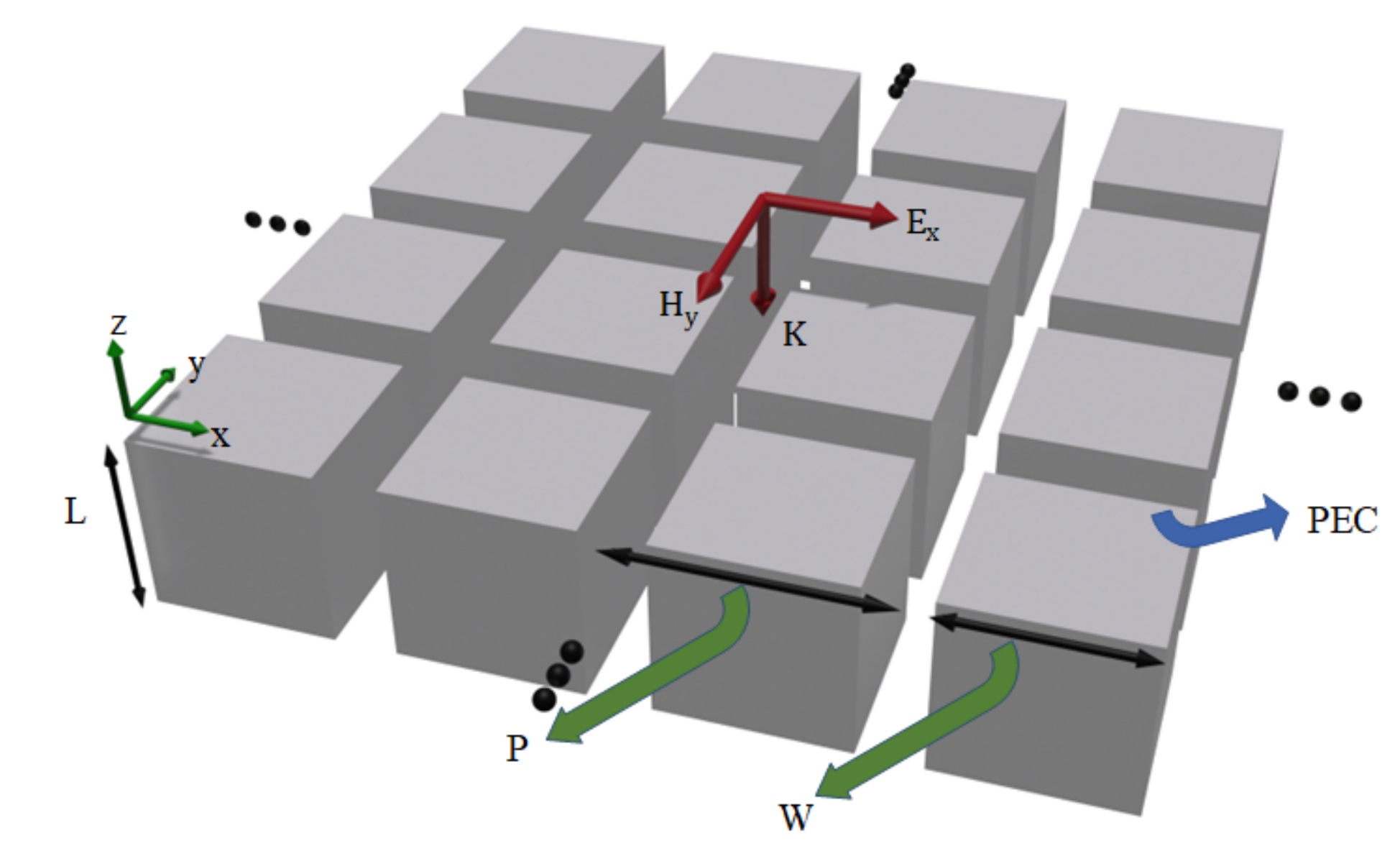}
	\caption{A periodic array of square PEC pillars with period P. The width of the pillars is W and the thickness of the pillars is L.\label{PEC_pillar}}
\end{figure}

In general, to find the realistic dominant mode profile inside the hole, we can increase the thickness of the array in full-wave simulations, and the electric field profile in the middle of the thickness can be used. For apertures with shapes such as circles, squares, and rectangles, the dominant mode has an analytical closed-form expression; however, a simulation profile can be used for unusual shapes. This case is suitable for arrays with a high thickness (almost thickness greater than $\frac{P}{4}$).

\begin{figure}[!ht]
	\centering
	\includegraphics[width=7cm,height=5cm]{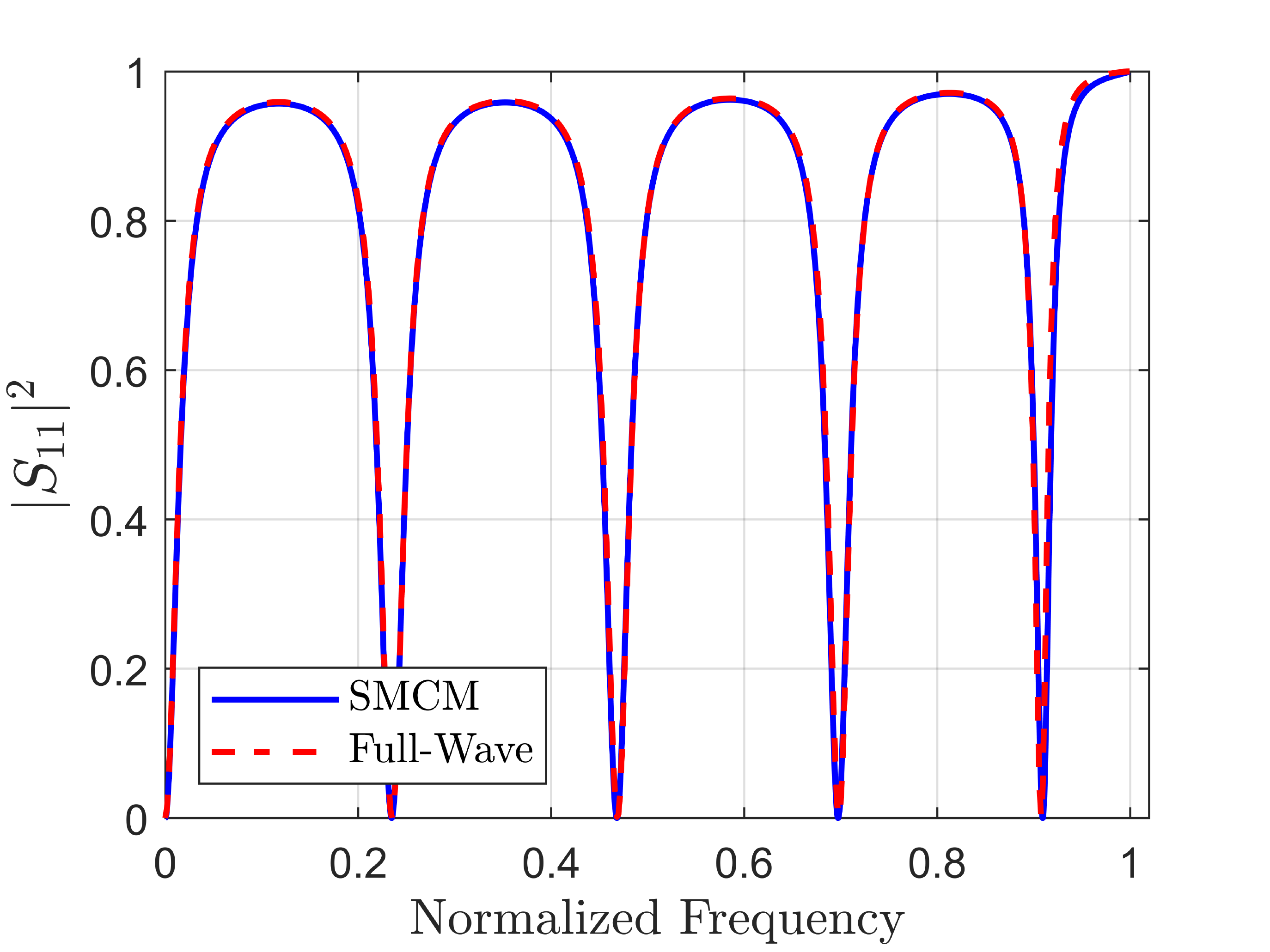}
	\caption{The result obtained from the single-mode circuit model for the array of PEC pillars (geometrical parameters: $W=.9P$,$L=2P$). The blue curve is the result of the circuit model, and the red dashed curve is the full-wave simulation result.\label{pill_results}}
\end{figure}

One of the structures that can be examined with the help of this model is the array of PEC pillars (Fig. \ref{PEC_pillar}). For this structure, the dominant mode which propagates between the pillars was obtained analytically \cite{edalatipour2015physics}. Using this profile, isotropic/anisotropic environmental models were presented and also if this profile is used in the SMCM, it gives accurate results (Fig. \ref{pill_results}).

\subsection{Multi-Mode Circuit Model (MMCM)}
\begin{figure}[!ht]
	\centering
	\includegraphics[width=8cm,height=5.5cm]{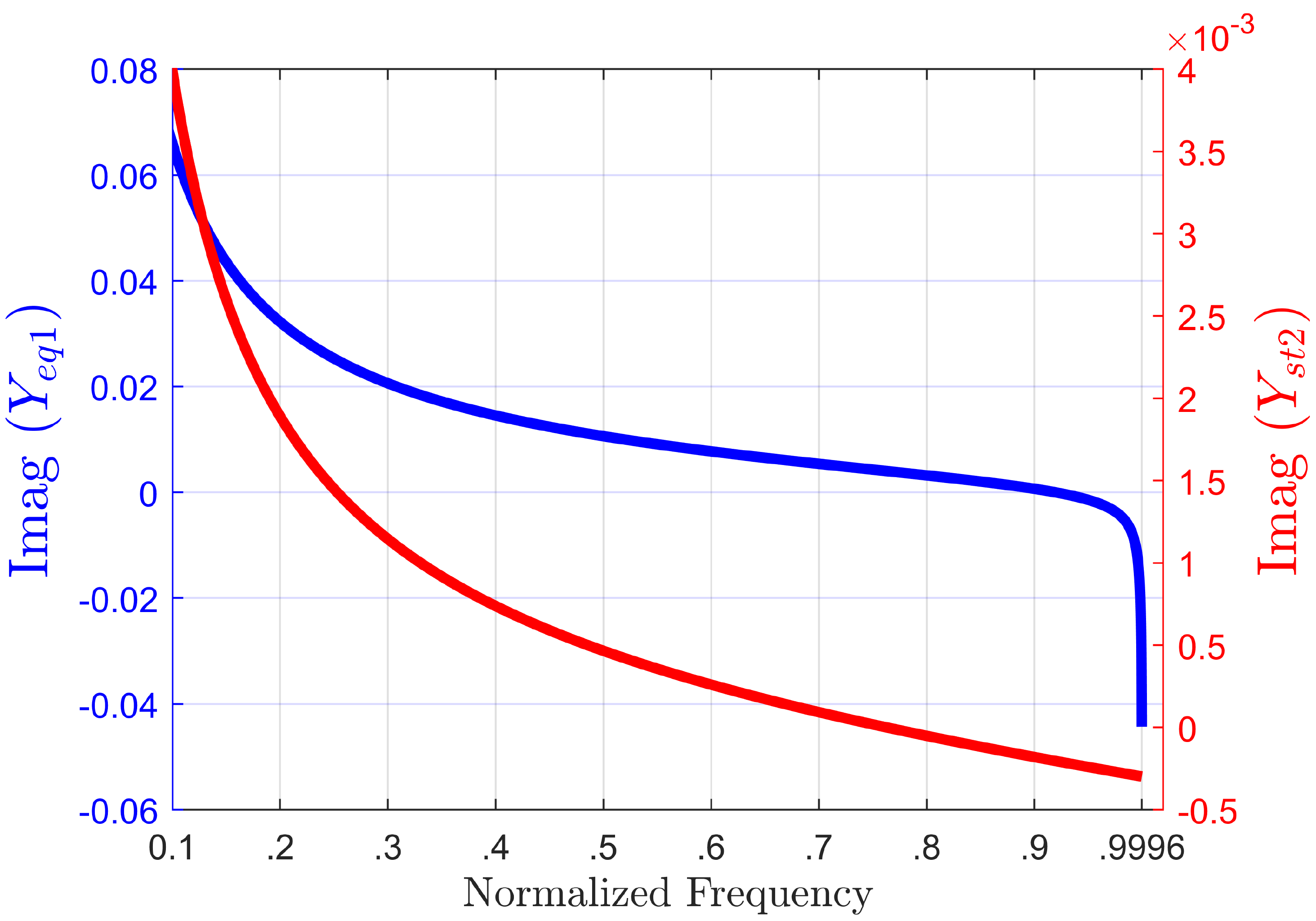}
	\caption{
		The red curve shows the admittance resulting from the modes inside the aperture except the dominant mode. The blue curve shows the admittance resulting from Floquet modes except the incident mode in region I.}\label{yeq_yst2}
\end{figure}

In the MMCM, we consider all the Floquet modes in regions I and III, as well as all the modes inside the apertures. In the MMCM, the electric field profile on aperture is not considered similar to the dominant mode and this profile is obtained from the full-wave simulator. First, let us examine an array of square holes in which the width of the squares is $\frac{P}{2}$, and the thickness of the holes is $\frac{P}{40}$. The electric field on the aperture at the resonance frequency obtained from the Comsol simulation is used to compute circuit model admittances (it should be noted that
we have used field profiles at the first resonance frequency). In Fig. \ref{yeq_yst2}, the values $Y_{st2}$ and $Y_{eq1}$ are depicted. TE and TM modes in the circuit model are considered inductors and capacitors, respectively. Due to the predominance of TE modes at low frequencies, they show an inductive effect, which gradually decreases with increasing frequency, and eventually, the capacitive effect will be dominant (the amount of transmitted power through the array is shown in Fig. \ref{cm_1}).

The MMCM can be more efficient in multilayer structures. First, we simulate the array at a frequency close to the resonance; then, with the help of the obtained profile, the multilayer structures with any number of layers can be examined. One of the disadvantages of the MMCM is that it requires an accurate electric field profile on the aperture, so we must obtain the field profile at a frequency close to the resonance frequency from the full-wave simulator and use it to compute admittances (of course, we can use the electric field profile at an arbitrary frequency, but the closer to the frequencies of interest e.g. resonance, the more accurate the results). Another complexity of the MMCM is that it is difficult to obtain high-order modes for unusual shapes; thus, we present another case that efficiently solves this problem.

\subsection{Modified Single-Mode Circuit Model (MSMCM)}
\begin{figure}[ht]
	\centering
	\includegraphics[width=8cm,height=5.5cm]{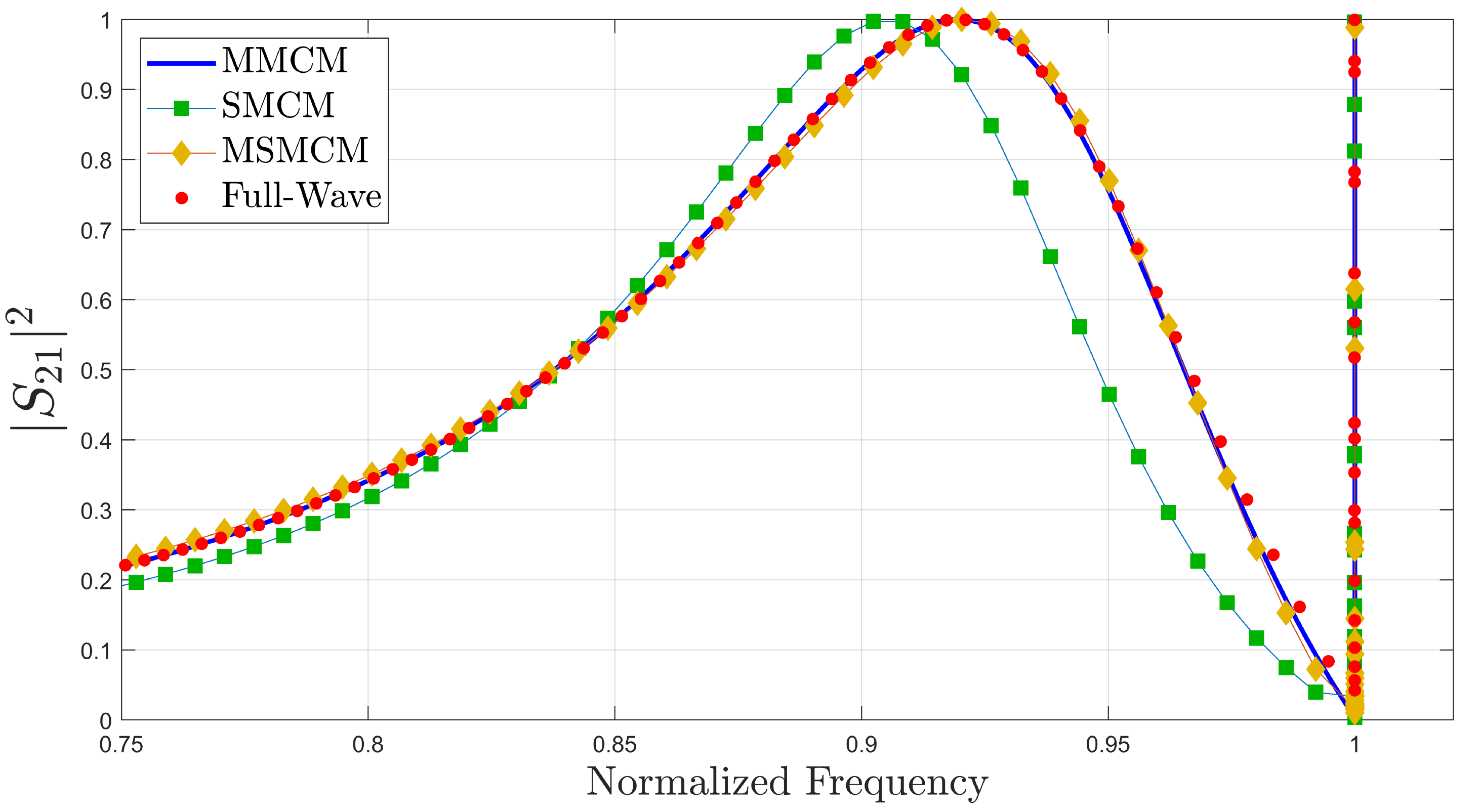}
	\caption{Comparison of the results of circuit models and full-wave simulation (parameters are $W=\frac{P}{2}$, $L=\frac{P}{40}$, $n_{1}=1$, $n_{2}=1$ and $n_{3}=1$).}
	\label{cm_1}
\end{figure}

The MSMCM similar to MMCM and SMCM considers all the Floquet modes in regions I and III, but considers only the dominant mode inside the aperture. The difference between the MSMCM and the SMCM is in the electric field profile. In the SMCM, the electric field profile at aperture opening is considered equivalent to the electric field profile of the dominant mode; However, in the MSMCM, we use the electric field profile on aperture obtained from the full-wave simulator (in the next section, for circular and square holes, closed-form expressions are presented). The modified single-mode circuit model gives results obtained in Fig. \ref{cm_1} for the array of square holes which was examined in the previous subsection.

As clearly seen in Fig. \ref{cm_1}, the circuit models predict extraordinary transmission (EoT). As can be seen, the accuracy of the MSMCM is less than the MMCM, but it is more accurate than SMCM. In the MSMCM, modifying the aperture profile has changed $Y_{eq1}$ and $Y_{st1}$ compared to the SMCM, which has significantly increased the accuracy. Since the MSMCM uses only one dominant mode, which makes this model more flexible than the MMCM, so we use this model in the following sections.

\section{Thin arrays of sub-wavelength apertures}
Due to the importance and applications of arrays of sub-wavelength apertures perforated in a thin film, we focus on these structures in this section. In the previous section, we saw that the two cases MMCM and MSMCM have high accuracy. In these models, we need to use the aperture field profile obtained from the full-wave simulation. If the structure's thickness is zero, closed-form expressions are presented as electric field profiles of both circle and square shapes; however, these profiles also produce good results for non-zero thickness arrays. It should be noted that in order to use the MMCM, we must have an accurate electric field profile on the aperture; therefore, we use the approximated profiles in the MSMCM.

\subsection{Circular aperture}
Apertures with different shapes have different properties. Changing the shape and size of the aperture can help us to control electromagnetic waves. Experimental results have been investigated for holes with different shapes, one of which is circular holes.
A closed-form expression has not been provided for the electric field profile of circular holes. Equation \ref{app_cr} is an excellent approximation of the electric field profile for circular apertures. 

\begin{equation}\label{app_cr}
	\begin{cases}
		E_{xa}=.1936\frac{1}{\sqrt{1 - 3.85(\frac{r}{D})^2}} \frac{\cos(\frac{\pi r\sin(\phi)}{D})}{\sqrt{1 - 1.12(\frac{r\sin(\phi)}{D})^2}}
		\vspace{.5cm} \\
		E_{ya}=.1074\frac{(\frac{2r}{D})^2}{\sqrt{1 - 3.68(\frac{r}{D})^2}} \sin(2\phi)
	\end{cases}
\end{equation}

\begin{figure}[!ht]
	\centering
	\includegraphics[width=8cm,height=7cm]{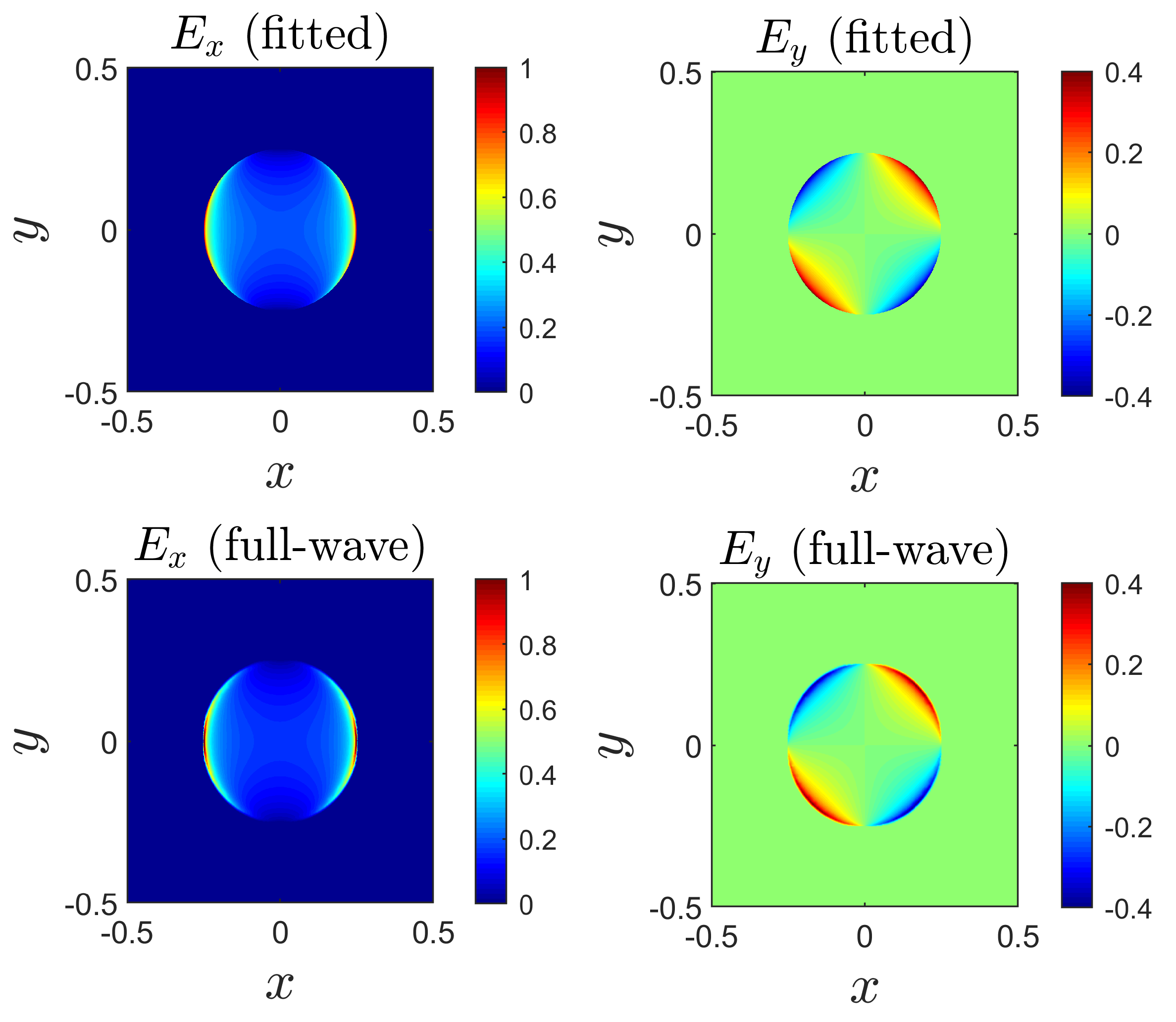}
	\caption{Transverse electric field profiles obtained from the full-wave simulation and  Eq. (\ref{app_cr})}\label{pro_cr}
\end{figure}

This spatial profile is approximated when the array is illuminated by a normal incident wave and provides an appropriate response in the non-diffractive region as long as $D<0.9P$ and $L<0.1P$. The actual (top row) vs approximated (bottom row) profiles are depicted in Fig. \ref{pro_cr}. Compared to rectangular apertures, in circular apertures, changes in the angle of incident waves have little effect on the reflection or transmission power; thus, these structures can be considered from this perspective. In the case of circular apertures, the dominant mode within the aperture is $TE_{11}$. 

\begin{figure}[!ht]
	\centering
	\includegraphics[width=17cm,height=6.5cm]{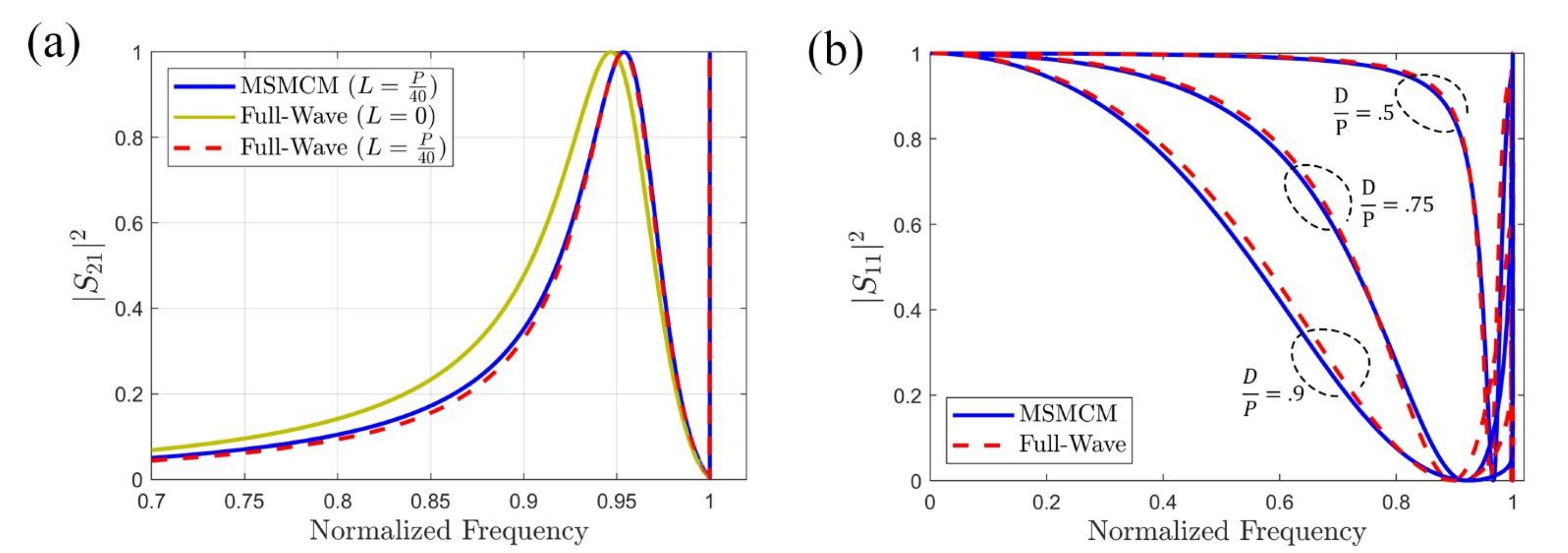}
	\caption{Figure (a) shows the results of full-wave simulations for two different thicknesses and the MSMCM result (parameters used in simulations are: $D=0.5P$, $n_{1}=1$, $n_{2}=1$ and $n_{3}=1$). Figure  (b) shows the comparison  between the circuit model and Full-Wave simulation results for different diameters (parameters used in simulations are: $L=\frac{P}{10}$, $n_{1}=1$, $n_{2}=1$ and $n_{3}=1$).}
	\label{cr_ll}
\end{figure}

Using the approximated profile for the circular aperture, the result in Fig. \ref{cr_ll} is obtained. As can be seen, the thickness $L=\frac{P}{40}$ in the full-wave results can make a significant difference at high frequencies compared to the zero thickness full-wave simulation. However the MSMCM has predicted the non-zero thickness result quite accurately, not previously possible with other methods. Since we use the MSMCM, the admittance resulting from the high-order modes inside the aperture is zero. If the thickness is zero, the middle transmission line resulting from the dominant mode inside the aperture is also removed.

\subsection{Square aperture}
\begin{figure}[!h]
	\centering
	\includegraphics[width=8cm,height=7cm]{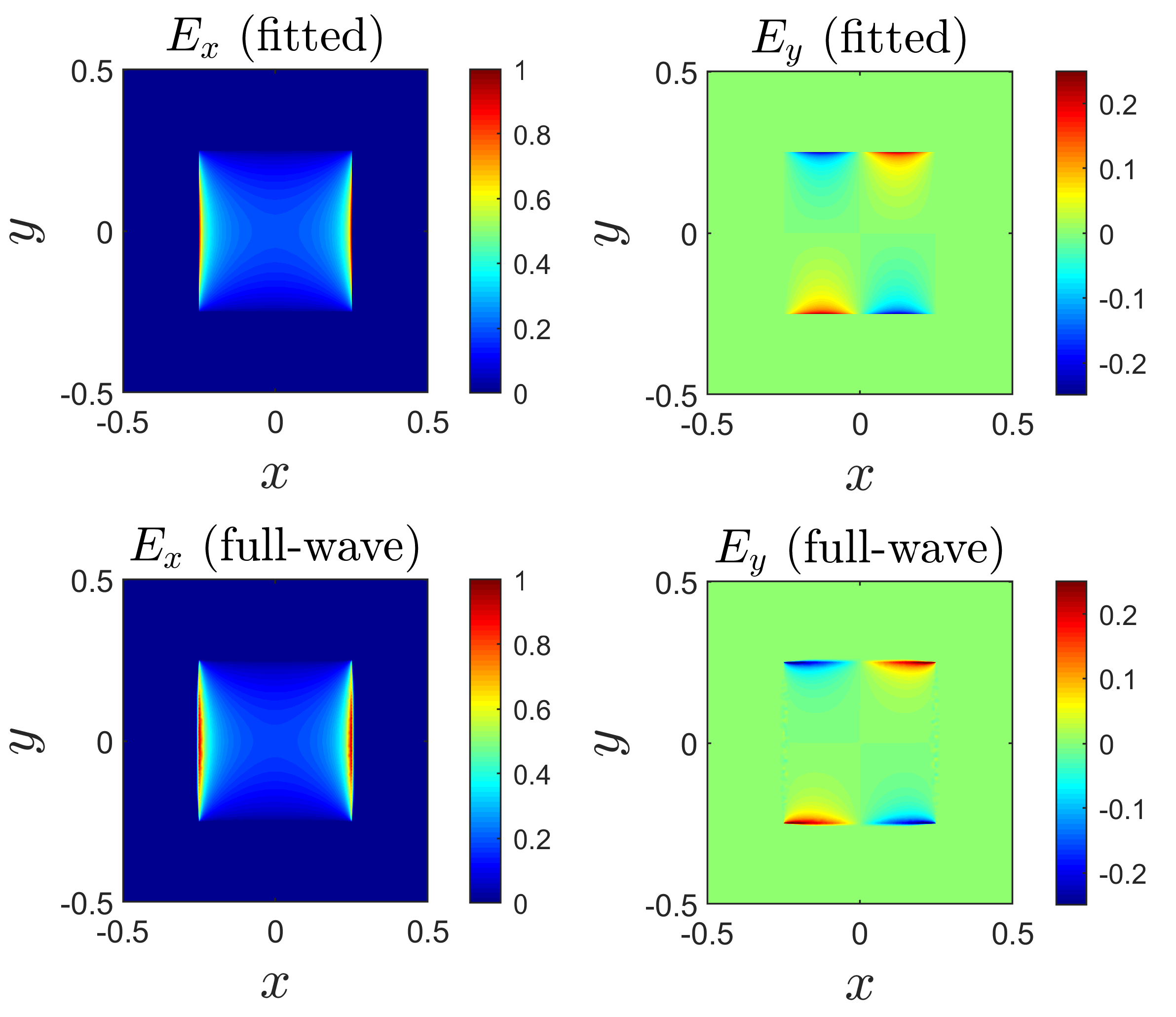}
	\caption{Transverse electric field profiles obtained from the full-wave simulation and Eq. (\ref{app_sq})} \label{pro_sq}
\end{figure}

\begin{figure}[!ht]
	\centering
	\includegraphics[width=17cm,height=6.5cm]{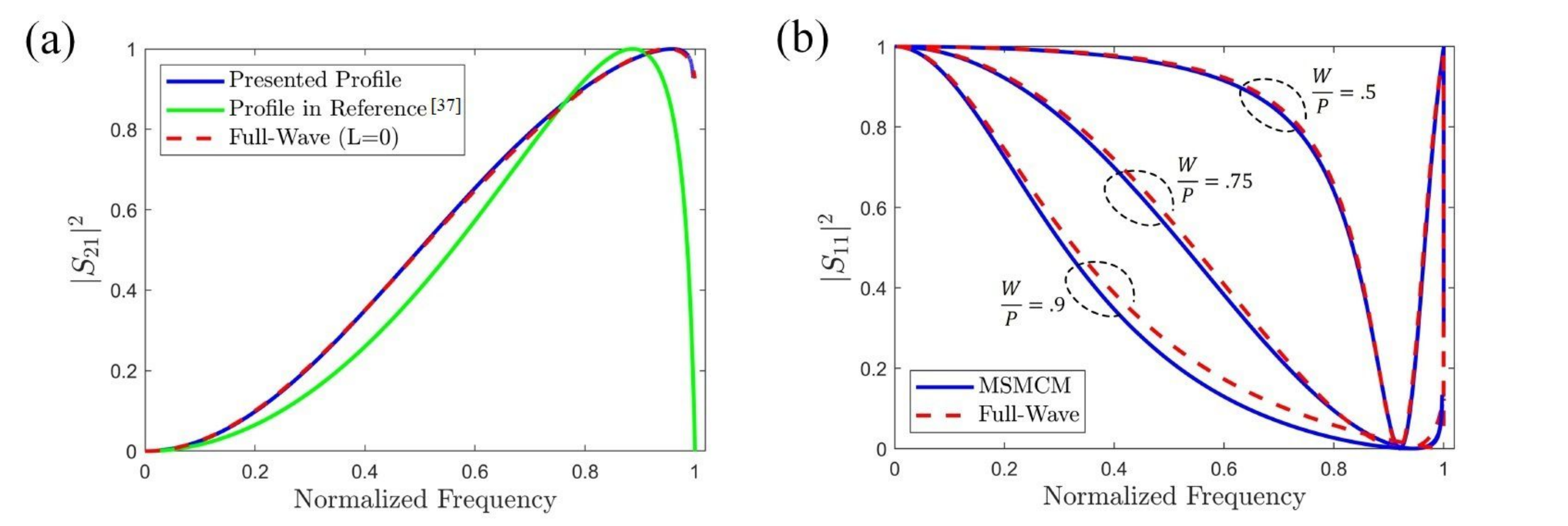}
	\caption{Figure (a) compares the results from the previous profile and the new profile ($L=0,W=.75P$,$n_{1}=1$ and $n_{3}=1$)). Figure  (b) shows the comparison  between the circuit model and Full-Wave simulation results for different widths (parameters used in simulations are: $L=\frac{P}{40}$, $n_{1}=1$, $n_{2}=1$ and $n_{3}=1$).}
	\label{sq_l0}
\end{figure}

Previous profiles for square shapes only consider field changes in one direction, which can cause errors, especially when the aperture is large (the width is more than half a period). The profiles used in previous studies produced accurate results for narrow rectangular apertures because the field changes along the breadth could be ignored. For square holes, we can use Eq. (\ref{app_sq}) as an electric field profile. Similar to the circular aperture, this profile also is approximated when the array is illuminated by a normal incident wave and provides an appropriate response in the non-diffractive region as long as $W<0.8P$ and $L<0.1P$. The comparison of the electric field profile obtained from the full-wave simulation and closed-form expression is provided in Fig. \ref{pro_sq} and using the approximated profile yields the results in Fig. \ref{sq_l0}.
\begin{equation}\label{app_sq}
	\begin{cases}
		E_{xa}=.1936\frac{\cos(\frac{\pi y}{W})}{\sqrt{1 - 4(\frac{y}{W})^2}} \frac{1}{\sqrt{1 - 3.85(\frac{x}{W})^2}}
		\vspace{.5cm} \\
		E_{ya}=.0928\frac{\frac{y}{W}}{\sqrt{1 - 3.85(\frac{y}{W})^2}} \sin(\frac{2\pi x}{W})
	\end{cases}
\end{equation}

\section{multilayer structure}
\begin{figure}[!ht]
	\centering
	\includegraphics[width=7cm,height=6.5cm]{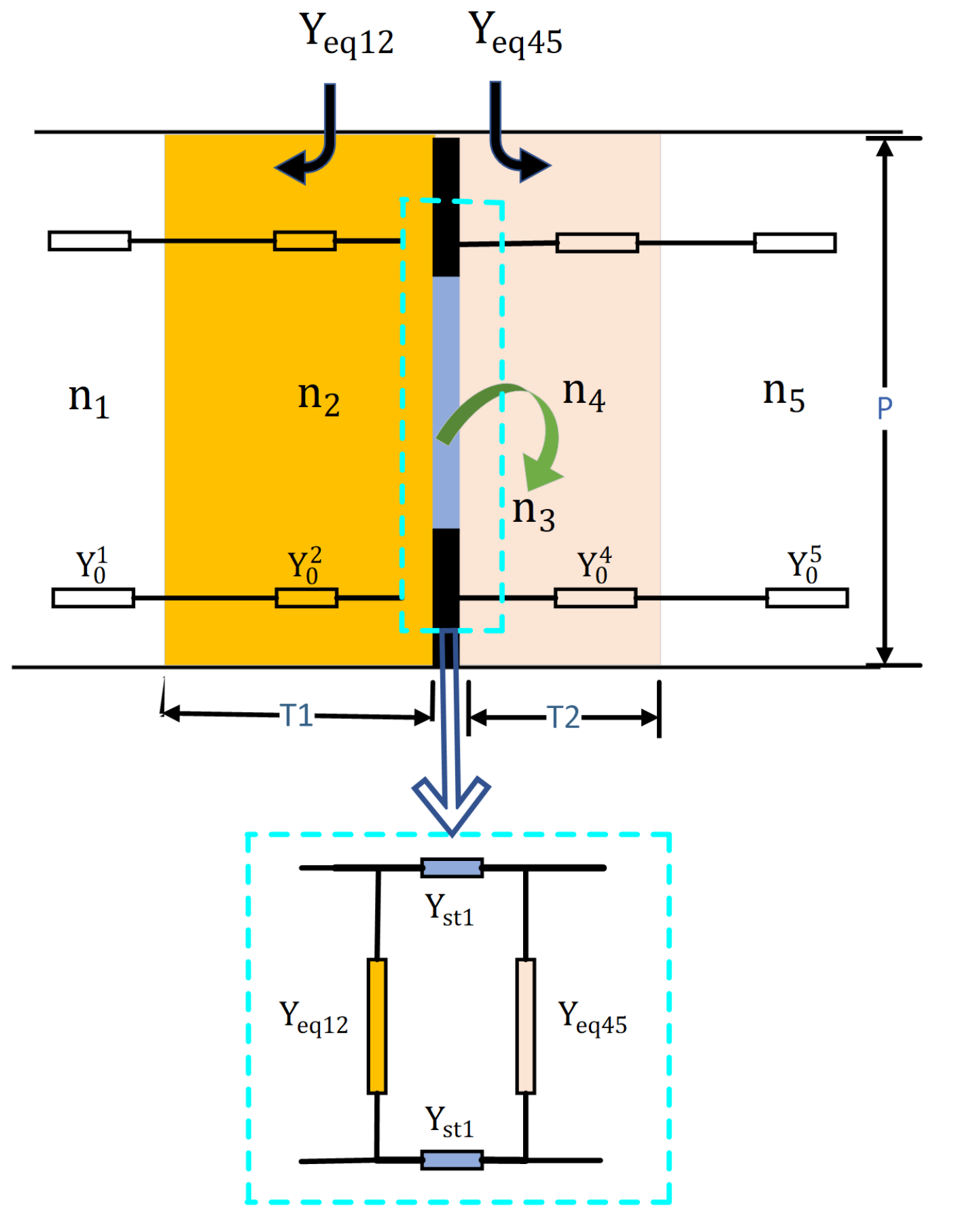}
	\caption{Modified single-mode circuit model for the array of circular apertures (Fig. \ref{periodic_array}(b)) with two layers added to the top and bottom. $T1$ and $T2$ represent the thickness of sample and substrate layers, respectively. The diameter and thickness of circular holes are $\frac{P}{40}$ and $D=\frac{P}{2}$, respectively.  \label{multi_layer_1}}
\end{figure}

Using an array of sub-wavelength apertures perforated in a film,  in practical applications requires placing the array on/between one or more dielectric layers. For example, to use these structures for applications such as sensors, typically a substrate is used, and eventually, this two-layer structure is placed on a layer of glass for testing \cite{yahiaoui2016terahertz,gordon2008new,dhawan2008plasmonic}. One of the advantages of the circuit model is its flexibility and simplicity in analyzing multilayer structures \cite{rodriguez2015analytical, alex2021exploring}. This feature further highlights the importance of developing accurate circuit models. In this section, we examine a biosensor structure and its sensitivity to the refractive index of the surrounding environment. For this purpose, we add two layers at the top and bottom to the array of circular apertures illustrated in Fig. \ref{periodic_array}(b). Figure \ref{multi_layer_1} shows the circuit model of this structure, where $n_{4}$ represents the substrate of this structure and $n_{2}$ refers to the sample under test. In this circuit model, we do not change the admittance related to the inside of the hole, however for the top and bottom layers, we have to modify the previous admittances as follows:
\begin{equation}
	\begin{cases}
		Y_{eq12}=\sum_{h}^{'}Y_{h}^{(2)} \frac{(Y_{h}^{(1)} - jY_{h}^{(2)}\tan(K_{zh}^{(2)}T1))}{(Y_{h}^{(2)} - jY_{h}^{(1)}\tan(K_{zh}^{(2)}T1))} \frac{A_{h}A_{h}^{*}C_{0}}{A_{0}A_{0}^{*}C_{h}}
		\vspace{.5cm} \\
		Y_{eq45}=\sum_{h}^{'} Y_{h}^{4}\frac{(Y_{h}^{5} - jY_{h}^{4}\tan(K_{zh}^{4}T2))}{(Y_{h}^{4} - jY_{h}^{5}\tan(K_{zh}^{4}T2))} \frac{A_{h}A_{h}^{*}C_{0}}{A_{0}A_{0}^{*}C_{h}}
	\end{cases}
\end{equation}

\begin{figure}[!ht]
	\centering
	\includegraphics[width=8.5cm,height=6cm]{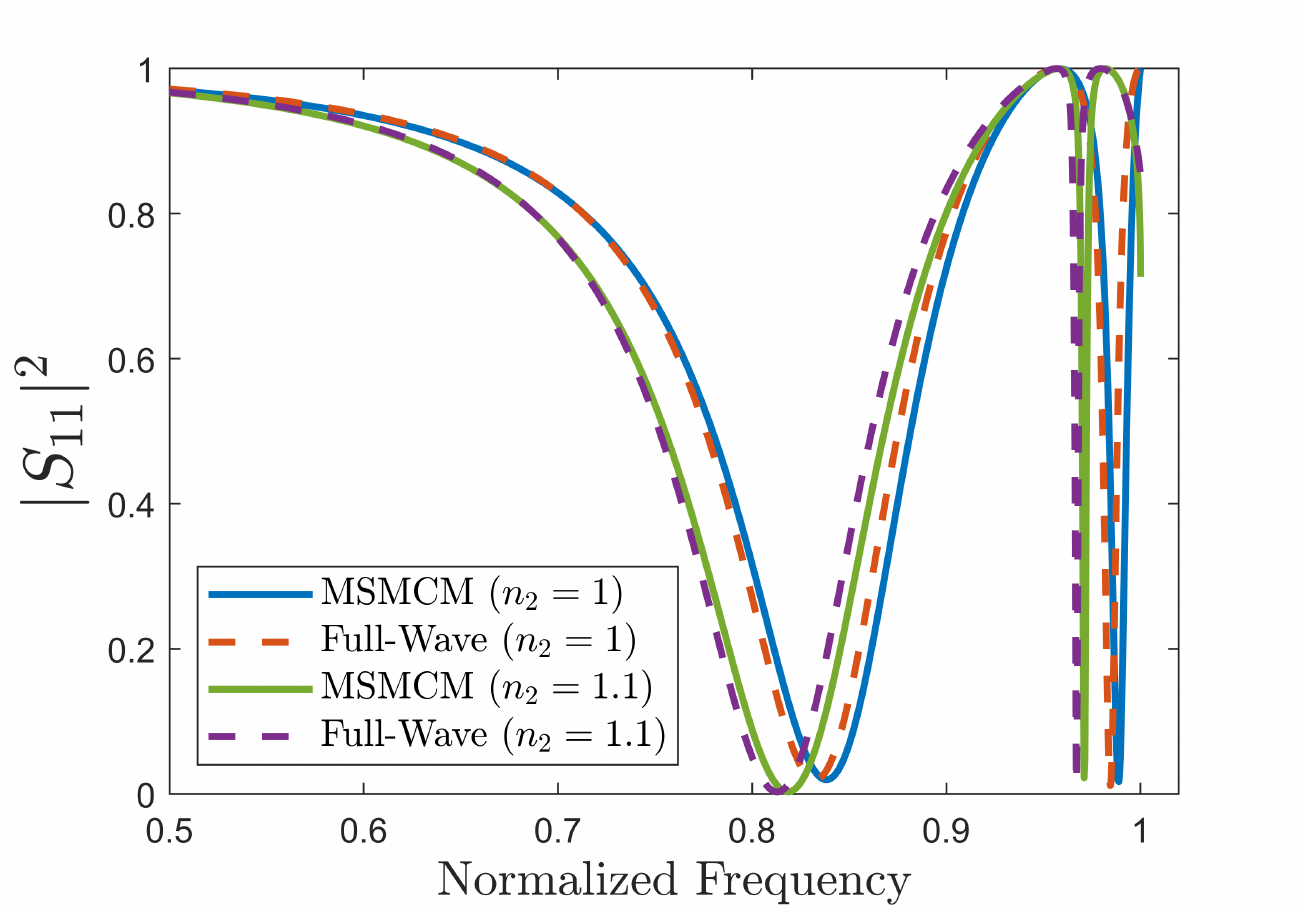}
	\caption{The reflection curve corresponding to the proposed sensor shown in Fig. \ref{multi_layer_1}. Parameters used in simulations are $T1=0.2P$, $T2=0.1P$, $n_{1}=1$, $n_{3}=2$, $n_{4}=2$ and $n_{5}=1$. \label{mult1}}
\end{figure}

With the help of the developed circuit model, the reflection of the biosensor has been computed in two cases where the refractive index of the sample is $n_{2}=1$ and $n_{2}=1.1$. As can be seen in Fig. \ref{mult1}, the circuit model's results have good agreement with full-wave simulation results and it has predicted the sensitivity of the first and second resonances with 99.7\% and 97.4\% accuracy, respectively.

\section{Conclusion}
In this paper, an extended circuit model was proposed for the arrays of sub-wavelength holes in a PEC film, by considering all the modes inside the aperture. The circuit model was examined in 3 general cases. The SMCM, due to the lack of need for electric fields profile at hole opening (dominant mode profile is considered as the profile at hole opening), is more useful than the others if the structure's thickness is large. If the thickness is large, for holes with common shapes such as circles, squares, and rectangles, the use of the SMCM is recommended. MMCM and MSMCM provide superior and accurate results, which is especially unprecedented for thin arrays. The MSMCM has higher performance than other models in thin arrays because it has high accuracy and, unlike the MMCM, does not need to obtain high-order modes inside the aperture, which means that the MSMCM has more flexibility. Using the approximated profiles and MSMCM, we obtained accurate results for thin arrays of circular and square holes.

\vspace{5mm}
	\textbf{Acknowledgements}
	
	The authors would like to sincerely thank Dr. Edalatipour for his assistance in preparing the paper.

\bibliographystyle{IEEEtran}
\bibliography{Reference.bib}

\end{document}